\documentclass[12pt,epsfig,citesort,svgnames,dvipsnames]{article}
\usepackage{multicol}
\usepackage[all]{xy}
\usepackage{graphicx}
\usepackage{todonotes}
\usepackage{amsmath}
\usepackage{amssymb}
\usepackage{amsfonts}
\usepackage{amsthm}

\theoremstyle{definition}
\newtheorem{definition}{Definition}[section]
\theoremstyle{remark}

\usepackage{amscd}
\usepackage{latexsym}
\usepackage[official]{eurosym}
\usepackage[english]{babel}
\usepackage[T1]{fontenc}
\usepackage[utf8]{inputenc}
\usepackage{authblk}
\usepackage{natbib}
\usepackage{xcolor}
\usepackage{hyperref}
\usepackage[nottoc]{tocbibind}
\hypersetup{
colorlinks=true,
citecolor=WildStrawberry,
anchorcolor=Black,
filecolor=cyan,
menucolor=magenta,
runcolor=cyan,
linkcolor=RoyalBlue,
linktoc=page, 
urlcolor=TealBlue}
\usepackage{tikz-cd}

\title{\bf {A Proposal for a Metaphysics of Self-Subsisting Structures.\\ I. Classical Physics}}
\author[a]{Antonio Vassallo
\thanks{\href{mailto:antonio.vassallo1977@gmail.com}{antonio.vassallo1977@gmail.com}}}
\author[a,b]{Pedro Naranjo
\thanks{\href{mailto:pnpfisica@gmail.com}{pnpfisica@gmail.com}}}
\author[c]{Tim Koslowski
\thanks{\href{mailto:t.a.koslowski@gmail.com}{t.a.koslowski@gmail.com}}}
\affil[a]{\normalsize Faculty of Administration and Social Sciences, Warsaw University of Technology, Plac Politechniki 1, 00-661 Warsaw, Poland}
\affil[b]{\normalsize Faculty of Philosophy, University of Warsaw, Krakowskie Przedmie\'scie 3, 00-047 Warsaw, Poland}
\affil[c]{\normalsize University of W\"urzburg, Theoretical Physics II, Campus Hubland Nord, Emil-Hilb-Weg 22, 97074 W\"urzburg, Germany}

\date{}

\begin{document}
\maketitle
\begin{center}
Published in \emph{Synthese} \textbf{200}, 374 (2022). \url{https://doi.org/10.1007/s11229-022-03865-x} 
\end{center}
\vspace{0.5mm}
\pdfbookmark[1]{Abstract}{abstract}
\begin{abstract}
    We present a new metaphysical framework for physics that is conceptually clear, ontologically parsimonious, and empirically adequate. This framework relies on the notion of \emph{self-subsisting structure}, that is, a set of fundamental physical elements whose individuation and behavior are described in purely relational terms, without any need for a background spacetime. Although the specification of the fundamental elements of the ontology depends on the particular physical domain considered---and is thus susceptible to scientific progress---, the empirically successful structural features of the framework are preserved through theory change. The kinematics and dynamics of these self-subsisting structures are technically implemented using the theoretical framework of \emph{Pure Shape Dynamics}, which provides a completely relational physical description of a system in terms of the intrinsic geometry of a suitably defined space called \emph{shape space}.
\end{abstract}
\tableofcontents

\section{Introduction}\label{sec:int}
In this paper, we take the word ``physics'' very seriously. In fact, we take it literally as ``(the study of) natural things.'' This is because we believe that physics goes hand in hand with a simple thesis, namely, that there are mind-independent ``things'' making up the natural world, which we get to know by taking scientific claims at face value.\footnote{It may be argued that such a simple thesis is tripartite since it involves metaphysical, epistemic, and semantic connotations. See for example \citet[][\S 1.2]{664}, for a discussion of this threefold nature of scientific realism.} 

This realist understanding of physics---and science in general---is often criticized on historical grounds: Past theories were considered successful until eventually proven wrong, and we have no reason to exclude that the same fate awaits currently well-established theories. Hence, how can we take \emph{any} theory at face value? Here we do not want to enter this debate (see \citealp{674}, for an excellent overview of the subject), and we will endorse the standard structuralist response to this challenge firstly put forward in its modern form by \cite{676}. In a nutshell, structuralists accept that there is a change in the theoretical entities postulated by physical theories as physics develops (e.g., string theory is quite obviously not fundamentally about Newtonian material particles). However, they maintain that the way such entities are related is preserved through theory change, so past theories were successful insofar as they captured these structural aspects of the world.

At first sight, structuralism might be taken as \emph{just} accepting that true scientific claims provide us with knowledge about the structural aspects of the world. In this sense, structuralism would represent a merely epistemological refinement of scientific realism. A historically remarkable example of this epistemic attitude can be found in Henri Poincar\'e, who famously claimed that the empirical success of a physical theory relies on its capability to provide an accurate description of the relational aspects of reality. For him, it is these relations that we can epistemically access, whereas the true nature of the entities bearing these relations to each other we will never get to know (\citealp{675}, p.~15).

However, structural realism may also be taken as a metaphysical thesis: The ontology of the natural world \emph{consists} of physical structures (see \citealp{535}, for one of the first papers acknowledging the difference between \emph{epistemic} and \emph{ontic} structural realism). There are many ways to shape this ontic structural realist thesis. One of them may, for example, be to take Poincar\'e's intuition to its extreme ontological consequences. This is done by denying that theoretical entities directly refer to anything like individual entities: There are only relations ``all the way down'' (see \citealp{681}, for a presentation of this radical ontic thesis). There is also a more moderate ontic take on physical structures. According to moderate ontic structural realism, individual physical objects are to be taken ontologically on a par with the relations they instantiate, which means that neither of the two is reducible to the other (see \citealp{517}, for an articulation of this position). In the following, we will focus on structural realism as a metaphysical position instead of a purely epistemic one.

The discussion regarding structural realism in physics is usually carried out in the context of ``standard'' physical theories (classical mechanics, general relativity, and even quantum theory). However, to our knowledge, not much emphasis has been placed so far on putting forward a theoretical framework that patently encodes the core tenet of structuralism, i.e., the idea that a physical theory has to be taken seriously only insofar as its relational aspects are concerned. The aim of this paper is precisely to discuss a general theoretical framework that can encompass the empirically adequate structural aspects of classical and quantum physics and which may serve as a template to implement a quantum gravity theory. However, in order to keep the discussion at a reasonable length, the analysis will focus on classical (i.e., non-quantum) physics, leaving the discussion of the quantum case to a companion paper.


The first step in characterizing this theoretical framework is to recognize that this structuralist reading naturally fits in with a relationalist take on dynamics in the context of theories of motion: Simply speaking, theories like classical mechanics are not about individual objects located in a background spacetime but, instead, about a web of spatiotemporal relations instantiated by said objects. Recently, some extensive attempts have been made to recast physics in Leibnizian/Machian relational terms according to the program set out in \cite{83}. This brand of relationalism accords a privileged status to spatial relations and seeks to reduce time to a mere change in these spatial relations. This approach has the advantage of being ontologically parsimonious while preserving empirical adequacy (with this respect, see the discussion in \citealp{470,469}). The present paper will show how these Leibnizian/Machian ideas are essential in setting up a unified framework for physics, which fully captures a structuralist understanding of theory change.

The plan of the paper is the following. In \S \ref{sec:psd}, we will introduce the technical backbone of the theoretical framework, including its main principles and some concrete applications. In \S \ref{sec:osr}, we will elaborate on the kind of metaphysics that goes along with the formalism. We will show how the best understanding of the framework is given in terms of \emph{self-subsisting} structures, which are configurations of fundamental objects individuated by a set of fundamental, world-building relations. We will also point out that, even if the concrete characterization of these fundamental objects and relations depends on the particular physical domain considered, the ``gist'' of these structures is preserved throughout these domains (and, hence, through standard theory change), in line with scientific structuralism. \S \ref{sec:hume} will be devoted to providing a metaphysical story about how space, time, and other salient dynamical features of the standard physical description of the world can be recovered from an ontology of self-subsisting structures. Finally, in \S \ref{sec:disc}, we will put forward some preliminary reflections on how the framework presented may play out in the context of the quantum gravity program.

\section{Pure Shape Dynamics}\label{sec:psd}

\subsection{Motivation}
\label{motivation}

One of the primary motivations for pursuing a relational strategy in constructing physical theories is epistemic and amounts to the realization that every measurement of a physical magnitude boils down to the comparison of said magnitude with a chosen standard unit. Hence, only ratios of physical quantities carry objective information.\footnote{This comparativist attitude with respect to physical magnitudes is not immune to philosophical controversy (see \citealp{725,723,722}, for a recent example of the debate). Here we will take comparativism for granted, leaving a philosophical defense of our choice to future work.} This means that it makes no sense to speak of values---or change thereof---of a physical magnitude \emph{simpliciter}: This information becomes physically meaningful---indeed, empirically \emph{observable}---only in relation to something else. It does not take a giant conceptual leap to elevate such an epistemic consideration to a metaphysical guideline: There are no ``absolute'' physical magnitudes and, therefore, any empirically adequate ontology of the physical world should avoid elements whose variation makes no observable physical difference. Historically, the main targets of the relationalist despise for unobservable entities are absolute spatial and temporal structures: Think for example of Leibniz' famous arguments against Newtonian space and time (see, e.g., \citealp[][\S 5.1]{705}).

The modern relationalist way endorsed in this paper to formally implement the rejection of an external space in which material bodies are placed can be roughly summarized as follows (cf. \citealp[][\S 2.1]{390}): Take a spatial configuration of material bodies and ``quotient out'' all the degrees of freedom associated with its being embedded in such an external space. In the case of an external $3$-dimensional Euclidean space $\mathbb{R}^3$, this amounts to identifying all the configurations related by transformations that belong to the groups of rigid translations $\mathsf T$, rotations $\mathsf R$ and dilatations $\mathsf S$, which jointly define the so-called {\emph{similarity group}}. Thus, if $Q^N$ is the configuration space associated with the geometrical configuration of a $N$-body system in $\mathbb{R}^3$, with each of the bodies placed at one of the vertices of the associated $N$-gon, the only objective spatial information encoded in the configuration is captured by the \emph{shape} of the $N$-gon or, more precisely, by its conformal structure.

Mathematically, this means that the relevant configuration space is $Q^N_{\mathrm{ss}}:=Q^N/\mathsf T\mathsf R\mathsf S$, which is known as \emph{shape space}. This quotienting out procedure can be extended to more general cases---including dynamical geometry---provided that there be a suitable sense in which the redundant structure to be washed away has an associated symmetry group.\footnote{This may not be the case for structures that represent conditions fixed a priori in the theory such as, for example, having the structure of a Hausdorff space. We will gloss over these cases for simplicity's sake, given that they have no major impact on our analysis.} All shape spaces are instances of stratified manifolds (on this score see, e.g., \citealp{446}; see also \citealp[][chapters 6 and 7]{449}, for a thorough technical discussion of shape spaces). Throughout the discussion, we will denote shape spaces in general by $\mathfrak S$. It is important to reiterate that the conformal structure left after the quotienting out is performed represents \emph{all} the information needed to characterize a relational configuration, which means that facts about the identity of the bodies making up an $N$-gon are unimportant with this respect. In order to implement this fact formally, we should perform the quotienting out procedure also with respect to the permutation group.\footnote{To make a concrete case, the permutation invariant version of the configuration space of $N$ point-particles in Euclidean $3$-space---$\mathbb{R}^{3N}$---is the space of all $N$-element subsets of $\mathbb{R}^3$---which is symbolized by ${}^N\mathbb{R}^{3}$. It is easy to see that all possible permutations of particles in a fixed configuration represent the same $N$-element subsets of $\mathbb{R}^3$. Hence, reducing $\mathbb{R}^{3N}$ to ${}^N\mathbb{R}^{3}$ accounts for the fact that there is no label attached to the particles in a configuration so that they can be swapped without changing the physical information encoded in the configuration.} Historically, the search for a full implementation of this Leibnizian take on space led to \emph{Shape Dynamics} (SD; see \citealp{390} for a nice introduction to the subject, with an emphasis on conceptual matters, and \citealp{514}, for a pedagogical, yet comprehensive, account). In the context of SD, only the ratios of distances between bodies carry physical meaning, which is achieved exactly by demanding that angles be preserved under the quotienting out procedure.\footnote{We acknowledge the debate, especially in the theoretical physics community, over the physical significance that \emph{absolute} scale or size possess. In particular, a different implementation than ours of relational ideas in physics is exhibited in general relativity, which does \emph{not} perform the quotient by scaling transformations (see \citealp{135,136}, for thorough discussions, and \citealp{419,729}, for the first models exhibiting scale-invariance in particle dynamics and dynamical geometry, respectively).} 

In order to fully specify a dynamical description of the system, it is important to implement the temporal side of relationalism. A simple articulation of this approach can be found in the writings of Ernst Mach \citep{110}. According to Mach, time is not a physical structure whose existence is independent of the material happenings that make up the physical world (think of the Newtonian universal clock that ticks without relation to anything else in the universe); rather, it is the result of an act of abstraction over the ordered change inherent into said material happenings. Hence, any adequate relational theory of dynamics must dispense with the absolute and external metric structure that usually models Newtonian time. Consequently, it is not the case that relational dynamics unfolds in a space which is the Cartesian product of $\mathfrak{S}$ with the real line: In that case, the ordered change in the shape of a system as encoded in its associated curve $\gamma$ in shape space would be labeled by a unique and global parameter $t \in \mathbb R$. The standard formulation of SD put forward in \cite{712} replaces such privileged global parameter $t \in \mathbb R$ with an arbitrary dimensionless local parameter $t/t_0$---$t_0$ being some arbitrary initial time value. In this way, SD exhibits an evolution of the initial data that does not depend on the particular choice of time units (in \S\ref{comparison} we shall give a detailed analysis of how to get rid of external temporal structures in dynamical geometry as well). However, once such a choice is made, the units of time remain fixed throughout the evolution, thus introducing a \emph{preferred} parametrization of the dynamical system in terms of the dimensionless parameter $t/t_0$. Hence, it seems that a reference structure is introduced in the dynamical description of the system, which is independent of its relational details. This goes against Mach's intuition that temporal facts should completely depend on facts about intrinsic changes in the system. Indeed, a truly relational theory requires that \emph{nothing} over and above the structure of shape space be needed to account for the evolution of physical systems.



Therefore, in order to make a step forward in the construction of a fully relational framework, it is crucial to get rid of the reference structures that appear in the definition of units of (local) increments of time. This can be done by banishing any concept whatsoever of parametrization of the dynamical curve $\gamma$, hence taking it as an \emph{unparametrized} curve $\gamma_0$.\footnote{An unparametrized curve $\gamma_0$ is an equivalence class $[\gamma]$ of paths $\gamma_i$ in the following sense: two paths $\gamma_1:\ell _1\rightarrow X,\, \gamma_2:\ell _2 \rightarrow X $ are equivalent if there is a non-decreasing, continuous map $\phi:\ell _1 \rightarrow \ell_2 $ such that $\gamma_1=\gamma_2\circ  \phi $. Here, $\ell_1,\ell_2$ are the domains of the paths (usually, $\ell _i\in\mathbb{R}$), $X$ is a set or some structure thereof, as a manifold (again, usually, $X\simeq\mathbb {R}^n$).} This is the essential idea underlying \emph{Pure Shape Dynamics} (PSD), which can be considered a natural evolution of SD (see \citealp{726}, for a thorough technical introduction to this new relational framework).

The major novelty of the relationalist picture advocated by PSD is exactly its insistence on using only \emph{intrinsic} geometric properties of $\gamma_0$ in $\mathfrak S$ in the description of the evolution of a given physical system, which is expressed in terms of the \emph{equation of state} of $\gamma_0$: A point $q^a\in\gamma_0$ corresponds to the full configuration of the system, which by construction is its shape \emph{qua} objective data, to which it is added the set \{$\alpha _I^a$\} of any intrinsic geometric properties of $\gamma_0$ needed to fully specify the evolution. The mathematical structure underlying this manifestly intrinsic nature of the description is the directional action of a local section $A(q^a,\alpha _I^a)$ in a suitable unit tangent bundle over $\mathfrak S$: $A(q^a,\alpha _I^a)=\frac{dq/ds}{d\alpha _I^a/ds}=\frac{dq}{d\alpha _I^a}$, i.e., the equation of state of $\gamma_0$ expresses the ratio of change of its intrinsic geometric degrees of freedom (see, again, \citealp{726} for the technical details).

Thus, the fundamental structure of $\mathfrak S$ describing the equation of state of $\gamma_0$---and, hence, the evolution of a physical system---is largely topological, which in turn guarantees the parametrization-irrelevant\footnote{The notion of parametrization irrelevance (i.e., no reliance on any parametrization whatsoever) was first introduced in \citet[][\S 1.4.1]{443} to make a distinction with parametrization invariance (i.e., no reliance on a fixed parameter).} nature of PSD even if some parametrization $s$ is used in computations. However, some minimal geometrical structure is certainly needed to describe the curve, namely a metric on shape space $g_{ab}$ (cf. the definitions of the unit tangent vector, \eqref{unittangent}, and of the curvature degree of freedom $\kappa$ introduced after \eqref{HamiltonianNbodyhomog}). Two facts are worth stressing, at this point: (i) A metric on shape space does \emph{not} measure size, but the amount of similarity in the configurations, and (ii) there exists a naturally induced metric on shape space,  the \emph{kinematic metric}, which is the metric entering the definition of the kinetic energy associated with a given physical system. We shall consider the explicit case of Euclidean geometry in \S \ref{sec:ephs}; in the case of dynamical geometry, the kinematic metric is the DeWitt supermetric (cf. \citealp[][in particular, footnote 11]{390}). 

Let us formally state the ideas above as the fundamental principle underlying PSD:

\theoremstyle{definition}
\begin{definition} The evolution of any physical system is uniquely determined by the equation of state of the intrinsic geometric properties \{$q^a,\alpha _I^a$\} of the \emph{pure}, i.e., unparametrized curve $\gamma_0$ in shape space $\mathfrak S$. This equation of state determines the ratios of change of \{$q^a,\alpha _I^a$\}.
\label{PSD}
\end{definition}


Thus, given a physical system, we shall express the equation of state of the unparametrized curve in its associated shape space as\footnote{\eqref{curve0} is the ``components'' version of the equation of state, which is more convenient to work with than its directional action counterpart.}

\begin{equation}
\begin{array}{rcl}
   dq^a&=&u^a(q^a,\alpha _I^a), \\
   d\alpha _I^a &=&\Omega _I^a(q^a,\alpha_I^a)\,,
   \end{array}
   \label{curve0}
\end{equation}

and demand that the right-hand side be described in terms of dimensionless and scale-invariant quantities, whose intrinsic change is obtained employing Hamilton's equations of motion. For consistency, the elements in $\alpha _I^a$ in \eqref{curve0} must exhaust the set of all possible dimensionless and scale-invariant quantities that can be formed out of the different parameters entering a given theory. Note also that dynamical parameters, such as mass, charge, and spin, enter the equation of state through \emph{scale-invariant} and \emph{dimensionless} quantities, as shall be illustrated in \S \ref{dynamics} and \S \ref{examples}. Hence, the conformal structure alone does not individuate any dimensionful quantity. Thus, it may be argued that dimensionful dynamical parameters and physical units are not needed by the theory and serve only as extra descriptive information about the physical system.

In \eqref{curve0}, $u^a$ is the unit tangent vector defined by the shape momenta $p_a$: 
\begin{equation}
    u^a\equiv g^{ab}(q)\frac{p_b}{\sqrt{g^{cd}p_cp_d}}\,,
    \label{unittangent}
\end{equation}
which allows us to define the direction $\phi ^A$\footnote{A direction is defined through $n-1$ components in an $n$-dimensional manifold, hence the difference in the label.} at $q^a$. It is through the unit tangent vector and the associated direction that the shape momenta enter Hamilton's equations, which are in turn used in the intermediary steps leading to the equation of state \eqref{curve0}. 


\subsection{Comparison with Shape Dynamics}
\label{comparison}

At this point, it is important to spell out in more detail the difference between PSD and SD. As we have already hinted, such a difference is crucial to achieving a better relational dynamics.

In a nutshell, the essential difference amounts to the fact that SD relies on the notion of parametrization of a curve to make sense of the dynamics, whereas PSD's fundamental physical description is given in unparametrized terms. This, of course, does not mean that dynamical curves in PSD cannot be described using a suitable parametrization, but bringing in such a piece of formalism is by no means necessary and just amounts to adopting a notational shortcut to simplify the calculations. In order to understand why the same cannot be said of SD, we should take a closer look at how the notion of parametrization has developed in the literature on SD.

The formulation of SD laid down in \citet{529} is motivated by the desire to avoid the ``many-fingered'' time featuring General Relativity (GR), i.e., to identify a \emph{global} variable that could serve as \emph{physical} time and, hence, give rise to physical evolution. This is achieved by considering a time-dependent Hamiltonian, which naturally induces a curve parametrization in terms of a genuinely \emph{physical} temporal parameter related to York time. Although the ``many-fingered'' time does not arise in particle models, it is nonetheless illuminating to run the argument keeping both particles and dynamical geometry, in order to stress their remarkable structural properties.

First, consider the following pairs of conjugate variables: (i) \{$\log R, D$\} in the particle model, with $R=\sqrt{I_{\mathrm{cm}}}$ (where $I_{\mathrm{cm}}$ is the center-of-mass moment of inertia) and $D=\sum _1^N \mathbf{r}_a^{\mathrm{cm}}\,\cdot \mathbf{p}^a_{\mathrm{cm}}$ is the dilatational momentum (where $\mathbf{r}_a^{\mathrm{cm}}$ and $\mathbf{p}^a_{\mathrm{cm}}$ are, respectively, the position and momentum of particle $a$ relative to the center of mass), and (ii) \{$V,Y$\} in dynamical geometry, where $V$ is the spatial volume and $Y$ is related to York time. As detailed in \citet{529}, a deparametrization procedure with respect to $D$ (respectively, $Y$) allows us to transform $\log R$ (resp., $V$) into a physical Hamiltonian and $D$ (resp., $Y$) into a monotonic \emph{physical} time variable. Importantly, $D$ (resp., $Y$) is an \emph{independent} variable, on top of the shape degrees of freedom.

Next, \citet{712,706} aim to eliminate these independent physical time variables from the theory: This is obtained through the introduction of a logarithmic time $\tau=\ln(|D/D_0|)$ (resp., $\lambda=\ln(|Y/Y_0|)$), which renders the Hamiltonian dimensionless and time-independent.

Crucially, although the time dependence has been eliminated through the logarithmic time, the fact remains that \emph{the parameter $\tau$ (resp., $\lambda$) is a necessary initial datum in the standard formulation of SD.} Hence, although there remains a freedom in choosing the origin of $\tau$ (resp., $\lambda$) by selecting a different value of $D_0$ (resp., $Y_0$), for two points $x, y$ in shape space the curves that pass through them with different values of $\Delta \tau:=\tau(y)-\tau(x)$ (resp., $\Delta \lambda$) do not coincide away from these two points. Therefore, the difference $\Delta \tau$ (resp., $\Delta \lambda$) is not just ``descriptive fluff'': It refers to a salient physical feature of the dynamics and, as such, should be taken ontologically seriously. In the end, it seems that SD is committed to the existence of external (i.e., non-shape) features of the physical world. This is what motivates the PSD program. As already emphasized, PSD's goal is to eschew \emph{all} external structures from the curve in shape space, whereby the dynamics is expressed in terms of the intrinsic, geometric properties of said curve---that is, solely in terms of shape degrees of freedom \{$q^a, \alpha _I^a$\} (as per principle \ref{PSD}, formalized by \eqref{curve0}). An \emph{independent} variable, however physical, is no exception. Thus, we demand that the curve in shape space be unparametrized.

The core difference between SD and PSD is best conveyed by considering how these two theories mathematically implement the main relational tenets. Standard SD satisfies the so-called \emph{Mach-Poincar\'e principle}, whereby a point in shape space and a tangent vector to it suffice to uniquely generate a $\tau$- (resp., $\lambda$-) parametrized curve in shape space. PSD, on the other hand, satisfies a modified version of this principle (enunciated in definition \ref{PSD}): A point and whatever number of \emph{higher}-order derivatives of the curve are needed to uniquely determine the curve in shape space. These higher-order derivatives are the elements of the set \{$\alpha _I^a$\}.

To make apparent how PSD improves upon SD, it is perhaps useful to distinguish two forms of relationalism: On the one hand, one can define that any ratio of sizes, even if compared at two distinct times, are relational quantities. This notion of relationalism does not introduce an absolute scale, but does require that an omniscient describer, as it were, of the universe be able to ``remember'' a previous size for comparison at a later time. On the other hand, one can define that only instantaneous ratios are relational, thereby not requiring that the omniscient describer of the universe possess the ability to ``remember'' a previous size for later comparison.

The relational meaning of $\Delta \tau$ (resp., $\Delta \lambda$) depends on which interpretation of relationalism is used: If one adopts the notion of relationalism that compares sizes at different instants of time, then $\Delta \tau$ (resp., $\Delta \lambda$) is the logarithmic ratio of expansion rates at the two instants. If one adopts the notion of relationalism that only shapes can be ``remembered,'' then $\Delta \tau$ (resp., $\Delta \lambda$) can be expressed in terms of the geometric data of the curve in shape space, so it is not an independent quantity in this case.

Thus, if one accepts that ``remembering'' sizes for later comparison is less relational than not being required to ``remember'' sizes at all---as we submit---then PSD is definitely more relational than the formulation of SD that requires the initial datum $\tau$ (resp., $\lambda$). In other words, the version of the Mach-Poincaré principle that PSD satisfies is more relational than the original version of it satisfied by standard SD, which means that PSD is an improvement on SD.


\subsection{Dynamics}
\label{dynamics}

One question of the utmost importance is how an unparametrized curve is to account for evolution. In standard relational accounts, a primitive notion of labeling is employed to describe the physical change of a system. The only property this label must meet is that it be monotonically increasing if it is to serve the role of a temporal bookkeeping device. Thus, by letting this label take on an ordered set of values, one gets the corresponding set of configurations of the physical system, the succession of which generates the associated curve in the relevant configuration space.

However, PSD features unparametrized curves, which means the above account of dynamics is not available. The challenge is hence to find an intrinsic feature of the system that serves the purpose of a physically meaningful labeling of change. Fortunately, PSD does in fact possess the resources to handle the dynamical evolution of a physical system. This is due to the insight---originally put forward in \cite{706}---that a shape contains structure encoded in stable records, whereby the evolution is towards configurations (i.e., shapes) that maximize the \emph{complexity} of the system. The result provides the desired ground for introducing a direction of change, which boils down to the direction of accumulation of the above-mentioned stable records.\footnote{Note that this idea is far from being uncontroversial. See, for example, the exchange between \cite{710} and \cite{711}.}

Given that the only fully worked-out physical system exhibiting generic formation of stable records is the $N$-body system, we shall use it as an example to motivate our approach to dynamics. It is worth pointing out that promising results come from the vacuum Bianchi IX cosmological model, where a natural candidate exists for a measure of shape complexity in this simplified model of dynamical geometry (see \citealp[][section 3.5]{712}). However, the extension of our arguments to full dynamical geometry and quantum mechanics is still a work in progress.

For current purposes, complexity is essentially the amount of clustering of a system, with a cluster being a set of particles that stay close relative to the extension of the total system. Next, we demand that the complexity function grow when (i) the number of clusters do, and (ii) the clusters become ever more pronounced, namely when the ratio between the extension of the clusters to the total extension of the system grows.

In the case of the $N$-body problem, a natural measure of the complexity of a system is:
\begin{equation}
    \mathsf{Com}(q)=-\frac{1}{m_{\mathrm{tot}}^{5/2}}\sqrt{I_{\mathrm{cm}}}\,V_N=\frac{\ell_{\mathrm{rms}}}{\ell_{\mathrm{mhl}}}\,.
    \label{complexity}
\end{equation}

In the above expression, $I_{\mathrm{cm}}$ is the center-of-mass moment of inertia:

\begin{equation}
    I_{\mathrm{cm}}=\sum _{a=1}^{N}m_a\,\mathbf{r}_a^{\mathrm{cm}}\cdot\mathbf{r}_a^{\mathrm{cm}}\equiv\sum _{a<b}\frac{m_a\,m_b}{m_{\mathrm{tot}}}r_{ab}^2:=m_{\mathrm{tot}}\,\ell^2_{\mathrm{rms}}\,,
\end{equation}

where $\mathbf{r}_a^{\mathrm{cm}}$ is the position of particle $a$ relative to the center of mass, $m_{\mathrm{tot}}=\sum _a m_a$, $r_{ab}=|\mathbf{r}_a-\mathbf{r}_b|$, and $V_N$ is the Newton potential:
\begin{equation}\label{newt}
    -V_{\mathrm{N}}=\sum _{a<b}\frac{m_a\,m_b}{r_{ab}}:=m_{\mathrm{tot}}^2\,\ell_{\mathrm{mhl}}^{-1}\,.
\end{equation}

In \eqref{complexity}, $\ell_{\mathrm{rms}}$ and $\ell_{\mathrm{mhl}}$ account for the greatest and least inter-particle separations, respectively. Thus, their ratio, \eqref{complexity}, measures the extent to which particles are clustered. In this simple case, it so happens that the opposite of \eqref{complexity} is none other than the shape space version, $C(q)$, of \eqref{newt}. Hence, $C(q)=-\mathsf{Com}(q)$ is, accordingly, referred to as the \emph{shape potential}, modulo the mass factor.\footnote{To be precise, all PSD models have an associated shape potential, which is related to the measure of complexity defined in each particular context.}

In \cite{706}, it is shown that, for $E_{\mathrm{cm}}\ge 0$, $I_{\mathrm{cm}}$ is concave upwards as a function of Newtonian time, and its time derivative (essentially, the dilatational momentum, $D$) is monotonic. This implies that $I_{\mathrm{cm}}$ is U-shaped, with a unique minimum at $D=0$ that divides all solutions in half. This is the so-called \emph{Janus Point}.\footnote{This designation was first put forward in \cite{709}.} Interestingly enough, the complexity function \eqref{complexity} has a minimum near this point and grows in either direction away from it, whereby automatically defining a time-\emph{asymmetric} dynamics for internal observers, i.e., ones within one of the two branches at either side of the Janus Point.


Moreover, it is a well-known result that the $N$-body system features generic solutions in which the original system splits into subsystems consisting of individual particles and clusters. Such almost isolated subsystems become increasingly isolated in the asymptotic regime \citep{717}, and will develop approximately conserved charges, namely the energy $E$, linear momentum $\bf P$, and angular momentum $\bf J$. These charges, in turn, enable us to define units $X$ of spatial scale and $T$ of time duration as:
\begin{equation}\label{equ:SubsystemUnits}
X^2=\frac{\bf J^2}{\bf P^2}\,,\,\,\textrm{ and }\,\,T^2=\frac{\bf J^2}{E^2}\,.
\end{equation}

This discussion naturally leads to two crucial remarks. First, the ever better isolated subsystems serve as local and \emph{stable} substructures, whose ever better conserved charges give rise to stable \emph{records}. This dynamically defines a direction of increasing complexity, measured by the complexity function \eqref{complexity}, which tends to grow secularly. We argue that this direction of increasing complexity should be identified with the so-called \emph{arrow of time} for internal observers. Thus, we arrive at a description of such an arrow of time in terms of purely intrinsic properties of the unparametrized curve in shape space (see \S \ref{sec:spst} for a more in-depth discussion of this point). Second, within the dynamically formed subsystems, there are pairs of particles that may function as physical rods and clocks, referred to as \emph{Kepler pairs}, due to their asymptotic dynamics tending to elliptical Keplerian motion.

A final remark is in place at this point. This treatment of temporal structures in terms of the complexity function is a common trait of PSD and SD---indeed, the above discussion was originally introduced in the context of SD. However, SD exhibits a patent redundancy in its description of temporal structures since it introduces the complexity function while retaining the \emph{physical} parameter $\Delta\tau$ (resp., $\Delta\lambda$). PSD, on the other hand, codes all reference to temporal structure in the complexity function alone, thus resulting conceptually cleaner.

\subsection{Examples}\label{examples}

In order to illustrate how PSD describes dynamical systems, in this subsection we will consider some physically interesting models and their associated equations of state of the respective unparametrized curves in shape space. Furthermore, this will explicitly show one of the central tenets of the program, namely, its use of only intrinsic geometric properties $\alpha _I^a$ of the curve. The discussion will also make manifest how the size of the set $\alpha _I^a$ depends on the complexity of the system to be described: The more complex the system, the more geometric properties are needed in its physical description. 

The examples below shall show the equations of state of a number of physical systems, but, in order not to obscure the presentation, we will not exhibit the mathematical structure underlying the manifest unparametrized nature of said equations (see \citealp{726}, for this construction for the $E=0\,\,\,N$-body system). Finally, a word is in order: As already mentioned above, the use of a parametrization is legitimate for the sake of computational ease, but we stress again that it does not belong to the fundamental structure of PSD itself.

\subsubsection{Geodesic System}

The dynamical system described by equation \eqref{curve0}, with the tangential direction $\phi ^A$ at the point $q^a$ as the only element in the set $\alpha _I^a$, corresponds to the simplest system in shape space, namely one whose curve is a geodesic, given by the Hamiltonian:


\begin{equation}
        H=\frac{1}{2}g^{ab}(q)p_a p_b\,,
        \label{Hamiltoniangeod}
    \end{equation}
where $g^{ab}$ is the kinematic metric. Absorbing the dimensionless mass ratios $\mu_i:=\frac{m_i}{M}$ into the configuration space metric (see \eqref{kmetric} below) allows us to set the overall mean mass $M:=\frac{1}{N}\sum_{i=1}^N m_i$ to unity. In order to work out the equation of state, let us conveniently use the arc-length parametrization of the curve with respect to $g_{ab}(q)$:

\begin{equation}
 \left(\frac{ds}{dt}\right)^2=g_{ab}(q)\frac{dq^a}{dt}\frac{dq^b}{dt}=g^{ab}(q)p_ap_b\,,
 \label{arc-length}
\end{equation}
where $s$ is the arc-length parameter, which readily yields

\begin{equation}\label{equ:KinematicEOM}
 \frac{dq^a}{ds}=g^{ab}(q)\frac{p_b}{\sqrt{g^{cd}(q)p_cp_d}}\,,
\end{equation}
where the right-hand side is the unit tangent vector $u^a$, \eqref{unittangent}, which, recall, defines the direction $\phi ^A$ at $q^a$. Taking any explicit functional expression $\Phi (q,p)$ for the direction, one gets:
\begin{equation*}
\frac{d\phi ^A}{ds} = \frac{\partial \Phi}{\partial q^a}\,\frac{dq^a}{ds} + \frac{\partial \Phi}{\partial p_a}\,\frac{dp_a}{ds}\,,
\end{equation*}
which, by means of \eqref{unittangent}, \eqref{arc-length} and \eqref{equ:KinematicEOM}, enables us to write the equation of state of the geodesic curve by means of the canonical equations of motion generated by the Hamiltonian \eqref{Hamiltoniangeod} as:

\begin{equation}
 \begin{array}{rcl}
   d\,q^a &=& u^a(q,\phi)\,,\\
   d\,\phi ^A&=& \frac{\partial \Phi}{\partial q^a}\,u^a(q,\phi)-\frac 1 2\frac{\partial \Phi}{\partial u^a} g^{bc}_{\phantom{bc},a}(q)u_b(q,\phi)u_{c}(q,\phi)\,,
 \end{array}
 \label{geodesic}
\end{equation}
where we have dropped the arc-length parameter $s$ to emphasize the underlying unparametrized nature of the curve.

An example of geodesic system is a universal configuration of classical non-interacting particles that indefinitely expands. Note how such a system, by construction, does not exhibit the formation of stable records.

\subsubsection{Newtonian $E=0\,\,\,N$-body System}\label{sec:ephs}
Let us consider the Hamiltonian with a generic potential $V$:
    \begin{equation}
        H=\frac{1}{2R^2}(D^2+g^{ab}(q)p_a p_b)+V(R,q)\,,
        \label{HamiltonianNbody}
    \end{equation}
where the overall mean mass $M:=\frac{1}{N}\sum_{i=1}^N m_i$ has been absorbed into the coupling constant of the potential, \{$q^a,p_a$\} are coordinates and momenta in shape space, and $D$ is the dilatational momentum. Moreover, $g^{ab}$ is the kinematic metric, which, in the case of Euclidean space, is given by the scale-free Euclidean metric on configuration space:
\begin{equation}\label{scalefac}
 ds^2=\frac{\sum_{I=1}^N\,d\bf r_I^2}{R^2}\,,
\end{equation}
where $R^2:=\sum_{I=1}^N\bf r_I^2$ denotes the square of the total scale in the center-of-mass frame. In ``global'' coordinates, the induced kinematic metric on shape space takes the explicit form:
\begin{equation}\label{kmetric}
 \frac{1}{R^2}\,g^{ab}(q):=
 g_{IJ}\frac{\partial q^a}{\partial r^I}\frac{\partial q^b}{\partial r^J}\,,
\end{equation} 
 where $r^I$ and $g_{IJ}$ ---which contains the dimensionless mass ratios $\mu_i:=\frac{m_i}{M}$ --- are some coordinates and metric, respectively, in configuration space. We shall focus on the case of a homogeneous potential, $V(R,q)=\beta R^k C(q)$, with $\beta$ an arbitrary coupling constant into which the overall mass $M$ above has been absorbed, which enables us to write \eqref{HamiltonianNbody} as an energy conservation constraint:
    
    \begin{equation}
        H=\frac{1}{2}(D^2+p^2)+\beta R^{k+2}C(q)=0\,,
        \label{HamiltonianNbodyhomog}
    \end{equation}
where $p\equiv \sqrt{g^{ab}p_a p_b}$ is the length of the shape momenta. Unlike the geodesic system \eqref{geodesic}, now one can build a further degree of freedom \{$\alpha _I^a$\}, namely $\kappa\equiv\frac{p^2}{\beta R^{k+2}}$, which is related to   $K^{-1}$, with $K$ the curvature of the curve, which, in turn, is given in terms of the acceleration vectors. Thus, $\kappa$ can be thought of as an \emph{intrinsic acceleration}, i.e., a measure of how much the curve traced out by a given physical system deviates from geodesic dynamics, \eqref{geodesic}. The intrinsic change of $\kappa$ is obtained, once again, by Hamilton equations of motion, yielding the following equation of state: 
    
    \begin{equation}
 \begin{array}{rcl}
  dq^a&=&g^{ab}(q)u_b(q,\phi)\\
  d\phi ^A&=&\frac{\partial\Phi_A}{\partial q^a}\,g^{ab}(q)u_b(q,\phi)-\frac{\partial \Phi_A}{\partial u_a}\left(\frac{1}{\kappa}\frac{\partial C(q)}{\partial q ^a}+\frac 1 2 {g^{bc}}_{,a}(q)\,u_b(q,\phi)u_c(q,\phi)\right)\\
  d\kappa&=&-(k+2)\kappa\varepsilon(q,\phi,\kappa)-2u^a(q,\phi)C_{,a}(q)\,,
 \end{array}
 \label{NbodyzeroE}
\end{equation}
where $\varepsilon\equiv\frac{D}{p}=\pm\sqrt{-\left(1+2\frac{C(q)}{\kappa}\right)}$ can be solved for in terms of \{$q^a,\phi ^A,\kappa$\} by means of the energy conservation constraint \eqref{HamiltonianNbodyhomog}. 

The immediate question that arises at this point is: How do we recover the standard Newtonian notions of scale and duration from \eqref{NbodyzeroE}? Although it is clear that it will be impossible to get absolute units of scale and duration, since the curve is void of any such structures, one can nonetheless meaningfully ask how definitions of absolute scales evolve. To explain the question, let us consider a curve $\gamma$ in shape space and two points $q^a$ and $q^b$ on it. Next, we will define the total scale $R$ of the system at point $q^b$ to be the unit of size $R_0$ and the total duration between $q^a$ and $q^b$ to be the unit of time $T$. Then, given a third point $q^c$ on $\gamma$, one may ask: what is the total scale $R$ measured in units of $R_0$? Likewise, what is the duration between $q^b$ and $q^c$ in units of $T$?. To answer these questions, we will give explicit equations for standard scale and duration in the case of the homogeneous system \eqref{HamiltonianNbodyhomog}. Following the suggestion in \citet[][\S 4]{135}, we will call these standards \emph{ephemeris scale} and \emph{ephemeris duration}. Historically, the ephemeris time was a duration standard used by astronomers and based on the intrinsic properties of the solar system (considered as a closed system). More precisely, the ephemeris time was the duration standard which made all the motions of the dynamically relevant bodies compatible with Newtonian dynamics (see also \citealp[][\S 13.2.4]{514}, for a discussion of ephemeris scale and time in the context of SD).

Using the arc-length parametrization condition \eqref{arc-length}, we obtain the ephemeris scale equation (cf. \citealp[][\S 3.2]{726}, for the technical details of the derivation):
  \begin{equation}
  \frac{d}{d\,s}\log R=\frac{D}{p}=\pm\sqrt{-\left(1+2\frac{C(q)}{\kappa}\right)}\,.
  \label{ephemerisScale}
\end{equation}

Notice that the shape potential $C(q)$ in Newtonian gravity is negative definite, which must be taken into account in \eqref{ephemerisScale}.

Likewise, the ephemeris duration equation is derived from
\begin{equation}
  \frac{d}{d\,s}\log\left(\frac{ds}{dt}\right)=\frac{d}{d\,s}\log p= -\frac{2}{\kappa}u^a(q,\phi)\,C_{,a}(q)\,.
  \label{ephemerisDuration}
\end{equation}

The unit of time $T$ is obtained by integrating \eqref{ephemerisDuration} between configurations $q^a$ and $q^b$. 

As expected, the right-hand sides of both \eqref{ephemerisScale} and \eqref{ephemerisDuration} refer only to intrinsic properties of the unparametrized curve in shape space. In particular, the (rate of change of the) complexity function \eqref{complexity} gives rise to the standard, global notion of duration, exhibiting the general result of the emergence of an arrow of time in terms of the increase in complexity of the $N$-body system. It is important to point out that (i) the ephemeris equations are model-dependent and (ii) there exists a many-to-one correspondence between Newtonian models and equations of state of curves in shape space, so the ephemeris equations are in general not uniquely associated with an equation of state of the curve in shape space.

We should also emphasize one key fact: Although the complexity function, \eqref{complexity}, is, by construction, genuinely a shape quantity, it does \emph{not} belong to the fundamental geometric properties of the curve in shape space, for the simple reason that it functionally depends on---and hence is reducible to---more basic data, namely, inter-particle separations. 
    
\subsubsection{Newtonian $E>0\,\,\,N$-body System}

In this case, the potential reads $V(R,q)=\beta\,R^k C(q)-E$, with the non-vanishing energy breaking homogeneity, which implies that one can no longer solve the energy conservation constraint, $0=\frac 1 2(p^2+D^2)+\beta\,R^{k+2}\,C(q)-R^2\,E$, for $\varepsilon$ in terms of \{$q^a,\phi ^A,\kappa$\}. Thus, we have to augment the dynamical system \eqref{NbodyzeroE} with an equation of motion for $\varepsilon$ (which, again, is obtained by means of Hamilton equations). We obtain: 
\begin{equation}
\begin{array}{rcl}
   d\,q^a &=& u^a(q,\phi)\\
   d\,\phi ^A &=& \frac{\partial \Phi_A}{\partial q^a}\,u^a(q,\phi)-\frac{\partial \Phi_A}{\partial u^a}\left(\frac{C_{,a}(q)}{\kappa}+\frac 1 2 g^{bc}_{,a}(q)u_b(q,\phi)u_{c}(q,\phi)\right)\\
   d\,\kappa &=& -(k+2)\kappa\,\varepsilon-2\,C_{,a}(q)u^a(q,\phi)\\
   d\,\varepsilon &=& \varepsilon\frac{u^a\,C_{,a}(q)}{\kappa}-(\frac{C(q)}{\kappa}+2\sigma)\,,
 \end{array}
 \label{NbodyposE}
\end{equation}

where a further dimensionless ratio $\sigma\equiv\frac{R^2E}{p^2}$ can be formed out of the parameters of the model, and can be solved for by use of the energy conservation constraint as:
    \begin{equation}
  \sigma:=\frac{R^2\,E}{p^2}=\tfrac 1 2(1+\varepsilon^2)+\frac{C(q)}{\kappa}\,.
\end{equation}

\subsubsection{$E=0$ Bianchi IX cosmological model}

Finally, we shall consider the simplest non-trivial case of a relativistic cosmological model, the so-called \emph{Bianchi IX} or \emph{Mixmaster} model \citep{720}, whose relational version was first analyzed in \cite{718}. Consider the following Hamiltonian:
\begin{equation}
  \label{HamiltonBcl}
  H_{\rm eff}=p^2+\left(2\Lambda
  -\tfrac{3}{8}\tau ^2\right)v^2-
  v^{\frac{4}{3}}C(q)\,,
\end{equation}
where $p^a$ are the shape momenta canonically conjugate to $q^a$, $v$ is the spatial volume, $\tau$ its canonically conjugate variable, and $C(q)$ is the associated shape potential.\footnote{We emphasize again that there is a shape potential for each physical system, which is related to the measure of its complexity, with the relation between the two depending on each case. The shape potential in \eqref{HamiltonBcl} is \emph{not} (minus) the simple ratio \eqref{complexity} (\citealp[][\S 3.5]{712}).}
The curve on shape space traced out by the Bianchi IX Universe is described by the following equation of state:
\begin{align}
  dq^a&=u^a(q,\phi)\,,\nonumber \\
  d\phi ^A&=\tfrac{1}{2\kappa}\tfrac{\partial\Phi}{\partial u_a}\,C_{,a}(q)
    \,, \nonumber \\
  d\kappa&=u^a (q,\phi) C_{,a}(q)
  +\tfrac{1}{2}\sigma\sqrt{\kappa}\,, \label{Bianchicl} \\ \nonumber
  d\sigma&=\tfrac{1}{\sqrt{\kappa}}\left(\tfrac{1}{4}\sigma ^2+\tfrac{2}{3}C(q)\right)-2\,\varepsilon\sqrt{\kappa}
  \,,\nonumber
\end{align}
where $\kappa\equiv p^2\,v^{-\frac{4}{3}}$, $\sigma\equiv\tau\,v^{1/3}$ and $\varepsilon\equiv\frac{\Lambda\,v^2}{p^2}$. \\
Finally, because of the constraint on the effective Hamiltonian \eqref{HamiltonBcl}, $H_{\rm eff}\approx 0$, we can solve for the parameter $\varepsilon$, yielding
\begin{equation}
  \varepsilon =\tfrac{1}{2
  }\left[\tfrac{1}{\kappa}\left(C(q)+
    \tfrac{3}{8}\sigma ^2\right)-1\right]\,.
\end{equation}

The homogeneous nature of this cosmological model simplifies the equations for the direction and curvature degrees of freedom (vanishing derivatives with respect to $q^a$), but it requires nevertheless a further degree of freedom as compared to the Newtonian $E=0$ $N$-body system: $\sigma$ comes about as a consequence of the expansion of the Universe. By considering the transition from \eqref{geodesic}, through \eqref{NbodyzeroE} and \eqref{NbodyposE}, to \eqref{Bianchicl}, it is now manifest how the increase in physical complexity of the system analyzed corresponds to an increase in the number of geometric degrees of freedom required to describe its dynamics.

\section{From Shapes to Self-Subsisting Structures}\label{sec:osr}

The question that we are interested in investigating at this point is: What is the best metaphysical picture that goes along with the PSD framework? The philosophically-minded reader must have noticed that PSD comes with a conspicuous amount of metaphysics already built-in. Suffice it to recall the main tenet around which the framework is centered, that is, the denial of the existence of anything whose change makes no physical difference. This is a strong metaphysical assumption, which justifies definition \ref{PSD} and its technical implementation, as seen in the previous section. However, the dynamics encoded in \eqref{curve0} is not just informed by the metaphysical tenet mentioned before; indeed, it in turn strongly suggests some metaphysical morals, as we are going to argue in a moment. This, for us, is a clear signal that a fruitful conceptual analysis of the PSD framework should treat physical and metaphysical aspects on a par, with neither of the two being merely entailed in a strictly logical sense by the other. Such an approach is close to what is usually known as \emph{natural philosophy}. According to this doctrine, the physical and metaphysical pictures of a theoretical framework are conceptually interwoven (see, e.g., \citealp{536}, for a recent discussion of this philosophical approach). In this sense, answering the above question is not a matter of just ``reading off'' an ontology from the framework's formalism in a neo-positivist fashion (see \citealp{364}, for a presentation and defense of this latter stance). Therefore, in what follows, our metaphysical elaboration on the conceptual foundations of PSD should be considered as taking place in parallel with the technical development of the framework.

With this remark about philosophical methodology in place, let us now turn to what the most satisfactory answer is for us to the question asked above. As mentioned in \S \ref{sec:int}, the PSD framework is in principle able to encompass not only the Newtonian and general relativistic domains of physics---as sketched in the previous section---but also the quantum one (which, as said, is discussed in a companion paper). This may sound puzzling from a metaphysical standpoint, given that such a comprehensive class of physical models is usually cast in terms of wildly different types of theoretical entities, such as material particles, fields, wave functions, strings, \emph{et cetera}. Does this mean that, according to PSD, we should take these theoretical entities at face value all at once? To answer this question, we should recognize that ``taking at face value'' can be intended in two related but slightly different senses. First of all, we can take a theoretical entity at face value by accepting that it directly refers to something physical, i.e., inhabiting the physical world, rather than being just a descriptively useful piece of formalism. Thus, we may take classical particles at face value because we can reliably assign them an active causal role in determining some observable phenomena. However, we cannot do the same with the Hamiltonian function, which is, in fact, just a compact mathematical way to summarize the physical state of said particles (the standard source for a defense of this entity-based realism is \citealp{682}). A second, stronger sense in which we may take a theoretical entity at face value is by claiming that it is part of the fundamental furniture of the world, i.e., an element of the ontology of the theory, in case it is a fundamental theory. If something is fundamental, it is ``out there'' in the physical world, but the converse does not necessarily hold, whence the need for distinguishing between these two senses of ``taking at face value''.

Under the light of this distinction, it does not seem reasonable to take \emph{all} the theoretical entities associated with the various physical models encompassed by PSD as the fundamental furniture of the actual world. This is because, in this way, the ontology of PSD would become bloated and uninformative. Indeed, adopting such an all-encompassing horizontal fundamental ontology would \emph{a priori} bar the possibility to construct a tree of ontological dependencies among entities---with its most welcome baggage of explanatory power. Otherwise said, it seems that a metaphysics where---for example---the existence of particles depends on fields, and the existence of fields, in turn, depends on that of strings has much more explanatory power than a horizontal one where all those entities are given all at once as a brute fact of the matter.

However, there are serious reservations also in taking all of these theoretical entities at face value in the milder sense of ``being out there.'' Most importantly, this choice would obscure the fact that there has been a conceptual shift in the use of these theoretical terms that corresponds to the development of more and more empirically adequate theories in the history of physics. For example, the talk of gravitational action-at-a-distance between material particles in Newtonian mechanics has long been replaced by the general relativistic talk of local interactions mediated by a gravitational field. There is no doubt that general relativity does a better job than Newtonian gravity in capturing some crucial features of gravitational phenomena, and this justifies the conclusion that the picture of reality provided by general relativity is more accurate than that provided by Newton's theory. It would then be awkward to claim that the pictures provided by the two theories coexist in the actual world.

A moment of reflection shows, however, that the PSD framework is not about theoretical entities \emph{per se}, but about the way such entities are interrelated. This is perhaps the most straightforward metaphysical moral associated with the quotienting out procedure introduced in the previous section. In other words, such a procedure does not determine the fundamental elements of reality \emph{as individual entities}, but just the \emph{relational properties} of configurations of them, i.e., the shapes properly said. Indeed the primary result of the quotienting out procedure is the ``translation'' of the physical system under scrutiny from standard configuration space to shape space. Nevertheless, shape space itself is just a (stratified) manifold whose points do not carry any information regarding how the standard configuration space was constructed in the first place. Instead, what shape space is sensitive to is how the quotienting out procedure singles out the physical relations making up a shape. This information is encoded in the very geometric structure of shape space itself. Thus, the shape space corresponding to a Newtonian $N$-body system is geometrically different from one corresponding to, say, Bianchi IX cosmology. This implies that, whatever the physical domain, PSD remains invariably a theory of \emph{shapes}---not just particles, or fields, or any other individual theoretical entity. This is indeed the rationale behind the claim that the dynamical law \eqref{curve0} represents a unified description of all of these physical domains. What \eqref{curve0} describes is the common behavior of the structural features of such domains. It is then easy to argue that the PSD framework naturally calls for a metaphysical interpretation in ontic structural realist terms.

We should be careful, however, in clarifying what type of ontic structuralism we have in mind. From what we have said so far, it may be inferred that we favor some sort of eliminativist type of structuralism, which does away with objects, and frames structures as clusters of relations \emph{simpliciter}. While this is a legitimate philosophical choice (but see \citealp{542}, for a critical voice), we are not sympathetic to it in this context. The reason for this skepticism is simple: Although the quotienting out procedure does not \emph{determine} the objects in a structure, it does not \emph{eliminate} them either. For any physical domain, the objects are ``already there'' before the procedure occurs, in the sense that their postulation is more or less implicit in the individuation of the degrees of freedom described by the standard configuration space. What the procedure does is to \emph{reconceptualize} the metaphysical role of these objects in the physical picture. Hence, for example, there is a clear sense in which PSD translates a Newtonian $N$-body system into a structure made up of material particles related through a web of Euclidean spatial relations. In other words, there is a clear sense in which shapes in PSD are implemented as concrete physical structures, as opposed to merely mathematical constructs. From this point of view, it is doubtful whether a merely eliminativist take on structures in PSD would be able to account for such an implementation, at least without adding some additional interpretive baggage on top of this radical structural metaphysics (see \citealp[][\S 2]{702}, for an argument along these lines, and \citealp{407}, for a decently worked out eliminativist framework that seeks to overcome such an objection).


The version of ontic structuralism that we favor is a more moderate one that treats objects and relations as ontologically on a par. This choice is motivated by the fact that, as already pointed out, the way PSD shifts the focus from ordinary configuration space to shape space retains a robust notion of shape as a configuration of individual elements of reality, with such elements ``adding up'', once piece after the other, to make up a whole configuration. This is precisely in the spirit of the Leibnizian account of space as the ordering of coexisting things, i.e., space as a whole resulting from many local material facts put together.


The discussion so far seems to beg the question: At this point, why not going for a metaphysics of objects \emph{tout-court}? The answer to this question is relatively straightforward: Because the PSD framework dispenses altogether with any spatiotemporal background in which such objects could be placed and hence individuated. The only way objects can be individuated in such a framework is through the relations they instantiate. There is no other fundamental defining feature of such objects beyond their relational aspects, in particular no primitive intrinsic identity or distinctness (recall from \S \ref{sec:psd} that the characterization of a shape does not hinge on facts regarding the identity of the \emph{relata}). Because of this, the shapes portrayed by PSD are very peculiar structures, which we dub \emph{self-subsisting} to highlight the fact that no aspect of their existence depends on something external to them, such as a background spacetime in which they are embedded.

In conclusion, we submit that the best metaphysics that goes along with the physical description encoded in \eqref{curve0} is moderately structuralist and is based upon the notion of self-subsisting structure (which is a metaphysically refined version of the concept of shape). However, still, it is not clear whether, for us, \emph{all} possible self-subsisting structures are fundamental or just ``out there''. Indeed, when asked what it is that, say, both Newtonian gravitation and dynamical geometry get right under the light of the PSD framework, our answer would be: Some purely relational aspect of the gravitational interaction that is hence preserved in the passage from Newtonian shapes to dynamical geometry ones. However, are Newtonian self-subsisting structures---i.e., point-like particles related by Euclidean spatial relations---to be considered ontologically on a par with their dynamical geometry counterparts---i.e., some sort of field magnitudes related by pseudo-Riemannian spatial relations? Answering this in the positive would obviously reintroduce the problem of mixing up radically different and perhaps incompatible pictures of the physical world.

To this challenge, we reply that there are two ways to address metaphysical questions in the PSD framework. The first is to do that in the context of a specific model. For example, we can consider the $N$-body Newtonian model as a faithful depiction of a world where fundamentally there are point-particles instantiating a web of Euclidean spatial relations. However, we perfectly know that such a world is not the actual one for the simple reason that such a fundamental ontology is not (entirely) empirically adequate. With this respect, the more empirically adequate the model, the more accurate the ontological picture will be (so, perhaps, by developing the PSD framework to include quantum-gravitational motions, we might get closer to the actual fundamental ontology of the physical world; see \S \ref{sec:disc}). There is no way in which we could take all the models as (fully) empirically adequate, so there is no question whether we can mix up radically different pictures of the actual world by going for this intra-model discourse. However, this does not mean that our metaphysics is strictly speaking model-dependent in a neo-positivist fashion. Indeed, there is a second sense in which we are straightforwardly realist towards all PSD models. As we have already pointed out, there are some purely structural inter-model aspects of PSD that correctly capture observable phenomena, and it is precisely these shared structural features that we consider as being ``out there''. Note that this is not just epistemic structural realism in disguise since there is in principle a class of models of PSD that are completely empirically adequate, and we take these (yet-to-be constructed) models to be the ones giving us the most accurate fundamental picture of reality.


\section{From Self-Subsisting Structures to Spacetime}\label{sec:hume}

\subsection{A Supervenience Basis for Laws and Dynamical Structure}\label{subsec:structure}

In order for the picture of the world in terms of self-subsisting structures to be viable, a story is required about how this relational metaphysics (i) accommodates the dynamical law \eqref{curve0}, and (ii) recovers the familiar understanding of material objects located in spacetime.

Section \ref{sec:psd} made it clear that \eqref{curve0} is a sort of ``law schema'' that encompasses the common structural features of different physical domains. The way \eqref{curve0} is then actually implemented depends on the details of the quotienting out procedure carried out on the starting non-relational theory. Hence, for example, from $N$-particle Newtonian mechanics equation \eqref{NbodyzeroE} is derived; likewise, equation \eqref{Bianchicl} follows from dynamical geometry, and so on. From a metaphysical point of view, the challenge is to provide a unified ``mechanism'' that links relational models to laws, thus making sense to consider the latter domain-related laws as occurrences of \eqref{curve0}. In light of this challenge, it seems that primitivist and governing accounts of laws are not well-suited for the framework. According to the former account (defended, e.g., in \citealp[][chapter 1]{130}), laws cannot be analyzed in more fundamental terms---they are brute facts of the matter. In this case, however, the structural resemblance of laws like \eqref{NbodyzeroE} and \eqref{Bianchicl} would remain unexplained and may very well be just a fortuitous fact. This would, in turn, demote PSD from an all-encompassing physical framework to a mere collection of models with no apparent (meta)physical connection. According to the governing view of laws (see, e.g., \citealp[][\S\S 9.3-9.6]{242}, for a critical discussion of this stance), instead, laws determine physical happenings (and not the other way round). In the PSD context, this view would imply that laws such as \eqref{NbodyzeroE} and \eqref{Bianchicl} would be external to self-subsisting structures, but that would bar such structures from grounding the structural resemblance between these laws. This weakening of the capability of self-subsisting structures to account for the unifying nature of \eqref{curve0} would go against the very spirit of PSD, which seeks to eschew anything external to shapes from the dynamical description. At this point, it seems clear that a better strategy to find a common metaphysical ``mechanism'' that leads from all the models of PSD to \eqref{curve0} is to adopt a non-primitivist and internal view of laws, i.e., one according to which it is physical happenings that determine the laws.


The first step to implement this strategy is to go back to the relationalist norm put forward in \S \ref{sec:psd}. Simply speaking, the norm states that commitment to any structure whose variation makes no observable difference should be avoided. The critical insight is to realize that this norm can be restated as a supervenience principle: We should be committed only to as much structure as that sufficient to constitute a complete, non-redundant supervenience basis for the dynamical laws of our framework. This provides us with the common metaphysical ``mechanism'' that explains the structural resemblance of laws like \eqref{NbodyzeroE} and \eqref{Bianchicl} and, hence, justifies their being considered occurrences of \eqref{curve0}. In short, the structural resemblance of, e.g., \eqref{NbodyzeroE} and \eqref{Bianchicl}, is inherited via a supervenience relation\footnote{Here we will gloss over the conceptual subtleties and the philosophical controversies related to the notion of supervenience (see, e.g., \citealp{kim}, for a thorough discussion of this concept).} from the (dynamical) structural resemblance of self-subsisting structures across the models of PSD. In this sense, it becomes clear why \eqref{curve0} captures the structural features common to all models of PSD: It is trivially \emph{determined} by such structural features.

The problem at this point is that there are many different choices of such supervenience basis that can recover \eqref{curve0}, each of which carries a different metaphysical flavor with it. Most importantly, one may or may not include in the basis some primitive modal facts in the guise of causal powers or dispositions borne by self-subsisting structures. In the first case, the supervening laws would inherit some weak governing connotations (see the review symposium \citealp{641}, for an exchange on this subject), while in the second case, they would play no determining role whatsoever. Hence, it is clear that the relationalist norm discussed above has to be supplemented with some further requirement to adjudicate between these two choices. With this respect, it seems natural to take ontological parsimony as such extra requirement: All things being equal, the best supervenience basis for \eqref{curve0} is the one containing the least amount of structure possible. Here we will take for granted that there is a consistent way to count the amount of structure in the ontology; see \citealp{436}, for a discussion on how to recognize and count physical structure. ``All things being equal'' translates into ``given a PSD model'', which clarifies that the type of ontological parsimony we have in mind is an intra-model one. This remark is important because, otherwise, it may be objected that the metaphysics we are proposing violates parsimony in that it regards a wide range of entities implementing self-subsisting structures at least as genuine physical possibilities. The fact that parsimony is used in this specific way poses no conceptual problem since considerations about simplicity crucially depend on the particular context considered (see \citealp{703}, for an in-depth articulation of this view). For example, one may demand a parsimonious fundamental ontology without posing any restriction on the type of ``higher-order'' entities that metaphysically depend on such an ontology (with this respect, see \citealp{704}). In fact, given that we are not interested in restricting the range of physical possibilities encompassed by PSD (on the contrary, we have made it clear that we consider this aspect as a virtue of the framework), we regard intra-model parsimony as the only simplicity requirement that counts in this context. This is because intra-model parsimony restricts the amount of fundamental structure \emph{at a world} (possible or actual).

This appeal to parsimony in the context of PSD is anything but arbitrary. Indeed, the framework (i) dispenses with external spatiotemporal structures and (ii) reduces in a precise mathematical sense the relevant dynamical features of a physical system to the geometric properties of a curve in shape space. Fact (i) is a clear sign that PSD favors an ontology with very thin spatiotemporal features (more on this in \S \ref{sec:spst}), but also fact (ii) points in the direction of ontological parsimony. To see this, consider that the geometry of shape space is not a ``stable'' feature of the formalism, i.e., its details depend on the particular system under scrutiny. This fact can be exploited (and will be exploited below) to argue that the shape geometric degrees of freedom that are not common to all the models of PSD are not part of the all-encompassing structural aspects of the framework itself---as encoded in \eqref{curve0}---and, hence, should not be taken as referring to features genuinely borne by self-subsisting structures. Instead, such model-dependent geometric features should be considered as a useful formal tool to describe in a succinct yet informative way how the dynamics of the self-subsisting structures unfolds in each specific case. If we buy into this line of argument, then it is easy to accept that the only fundamental stuff equation \eqref{curve0} describes is structures featuring a set of fundamental relations instantiated by otherwise featureless \emph{relata}. Once we embrace such a parsimonious metaphysics, the next step is to deny any need for the presence of genuine modal properties in the supervenience basis. These properties are usually individuated in terms of---or even identified with---the causal role they play (e.g., mass being the disposition to move in a certain way in a gravitational field), but their fundamental role becomes dubious in a metaphysics that denies any substantial ontological import to dynamical parameters such as mass, charge, and the like.


In the literature, it has long been recognized that an adequate ontology for physics can, in principle, do away with fundamental modal features in the guise of intrinsic causal properties borne by objects. This is, in a nutshell, the central tenet of regularity theories of laws of nature, commonly referred to as \emph{Humeanism}. One of the best worked out examples of such a stance is the Mill-Ramsey-Lewis Best System account of laws of nature. According to this account, the laws at a world $w$ supervene on the arrangement of local matters of particular fact making up $w$---the \emph{Humean mosaic}---as the axioms of the simplest and most informative deductive system from which true physical statements can be derived (see \citealp{510}, for an introduction to this topic). This stance has been recently further developed in order to accommodate the non-locality inherent into quantum entanglement. The crucial move with this respect is to recognize that the mosaic does not need to include any intrinsic property \emph{at all}---not even the perfectly natural and categorical properties originally postulated by David Lewis (on this score see, e.g., \citealp{382}, and \citealp{477}, but see also \citealp{606}, for a critical assessment of this move). So how exactly does such a propertyless Humean metaphysics apply to the case of the PSD framework?

To answer this question, let us first of all turn to the formal machinery introduced and discussed in \S \ref{sec:psd}, and try to pinpoint the minimal set of degrees of freedom that are strictly needed to construct the dynamics and are common to all PSD models. These elements can be easily individuated to be shape space points $q^a$ and the direction $\phi^A$ at $q^a$. Indeed, $q^a$ and $\phi^A$ enter the description of the simplest models of the framework---i.e., geodesic motions: There is no simpler, physically meaningful motion modeled with less degrees of freedom than this. On the other hand, more complex motions can be modeled by introducing further geometric degrees of freedom \{$\alpha_I^a$\} \emph{alongside} $q^a$ and $\phi^A$. Under our parsimonious reading, \{$\alpha_I^a$\} are just needed to \emph{describe} in detail such motions but, in the end, do not amount to adding anything metaphysically substantial on top of the fundamental degrees of freedom $\langle q^a, \phi^A \rangle$, which are the only variables common to all models of the framework---i.e., they are \emph{necessary} to make sense of the dynamics in the first place. Think, for example, of the curvature degree of freedom $\kappa$ that arises in the simplest non-geodesic case \eqref{NbodyzeroE}. This degree of freedom is not needed in the geodesic case \emph{and} it is constructed out of the kinematic metric $g_{ab}$, which measures the ``intrinsic'' differences between shapes. This suggests that facts about particle accelerations can be reduced to facts about how the spatial relations among these particles change. Another concrete example that motivates our restrictive realist attitude is given by \eqref{complexity}---the expression of the complexity function $\mathsf{Com}(q)$ in the Newtonian context---, which is entirely given in terms of (ratios of) inter-particle separations. This means that the notion of complexity is not primitive but supervenes on---indeed, it is reduced to---intrinsic facts about shapes. To sum up, we take the structure $\langle q^a, \phi^A \rangle$ to refer to the Humean mosaic on which the law \eqref{curve0} and its dynamical structure supervene.

This ``metaphysical'' focus on the minimal set of degrees of freedom is a quite natural choice as far as our realist and parsimonious attitude is concerned: Since the physical degrees of freedom are the subject of the dynamical description, they are the most plausible candidates in the PSD formalism to refer to the features of the physical world. At this point, one may wonder why the shape space metric $g_{ab}$ shouldn't be taken metaphysically seriously as well, given that it is required to define unit tangent vectors and extract the curvature from the acceleration vectors---and, thus, to generate the dynamics. To this we reply that, although $g_{ab}$ occurs in the right-hand sides of all equations of state given in \S\ref{examples}, there is no equation for $g_{ab}$ itself. Hence, $g_{ab}$ is not the subject of the dynamical description, but a formal tool used to facilitate such a description. Under this view, claiming that $g_{ab}$ represents a \emph{sui generis} feature of reality would be similar to reifying the Hamiltonian function---a rather awkward ontic commitment that would inflate the ontology without adding explanatory value.


The next step is now to characterize the entities in the mosaic. With this respect, we look at the minimal amount of facts needed to make sense of fundamental degrees of freedom $\langle q^a, \phi^A \rangle$. The first half of this structure is represented by the points $q^a$, which already have a fairly straightforward characterization put forward in the previous section. Each point represents a universal configuration of fundamental elements of reality in an intra-model sense---e.g., particles, field magnitudes. These fundamental elements of reality are connected through a web of relations that are spatial in nature, and which are described by the conformal structure that ``survives'' the quotienting out procedure on the redundant spatial degrees of freedom of the particular system considered. We now see why intrinsic properties are not needed at all in this context. Simply speaking, these properties do not enter in any way the characterization of the points $q^a$ representing our self-subsisting structures. Compare this to the status of the relations making up a self-subsisting structure, which instead \emph{determine} the set of points $q^a$ (i.e., shape space itself) through the quotienting out procedure discussed in \S \ref{sec:psd}. 

The second half of the structure $\langle q^a, \phi^A \rangle$ tells us how such self-subsisting structures are dynamically related. From the discussion of the dynamical mechanism underlying \eqref{curve0}, we already know that the fundamental ordering of points $q^a$ in a dynamical curve is not metrical in the usual sense of the word. That is, there is no fact of the matter about a configuration $q^a$ coming, say, $n$ seconds after or before another configuration $q^b$. However, this ordering is not the weaker one encoded in a parametrization of a curve either, which can be rendered in terms of an ``earlier than'' relation. Such a relation has, in fact, a global character that is absent in this context. Simply speaking, given any two points on the curve, it is impossible to say which one comes ``earlier'' without a parameter that assigns a monotonically increasing numerical flag to each of them. On the other hand, the fact that $\phi^A$ is defined in a neighborhood of $q^a$ and depends on the kinematic metric $g_{ab}$ (which measures how much two shapes differ from each other) through \eqref{unittangent} allows us to conceive of the fundamental dynamical ordering as just a minimal topological ordering where the notion of nearness of configurations is rendered in terms of a \emph{similarity} relation (intended as lack of distinctness; see \S \ref{sec:spst}). According to this topological ordering, the most we can say is whether, e.g., a configuration $q^b$ is ``in-between'' $q^a$ and $q^c$: This happens just in case all neighborhoods of $q^a$ that include $q^c$ also include $q^b$.

It is then clear that it is challenging to \emph{globally} characterize the ``timelike'' or dynamical part of the mosaic in terms of \emph{succession} or \emph{change} in the configuration. Our metaphysics simply does not allow for anything like that at the fundamental level, allowing only for a ``local'' counterpart of change. The way out of this impasse is, first of all, to remember that a model of PSD represents a possible world, and then to recognize that such a possible world features the physically realized configurations represented by the corresponding solution of \eqref{curve0} given \emph{all at once} in a timeless and (almost) changeless sense. It is worth reiterating that this does not imply that the set of physically realized configurations is entirely unordered. On the contrary, there is still a weak, local, topological ordering of such configurations that makes it possible for the mosaic to be described using an unparametrized curve in shape space. To sum up, our metaphysical take on the PSD framework is the following: The framework depicts a cluster of physically possible worlds (models), each of which consists of a Humean mosaic of self-subsisting structures ``timelike''-arranged in a weak topological ordering that can be described in the simplest and most informative way by an unparametrized curve in the corresponding shape space.

It is important to note at this point that, recently, a similar metaphysical take on physics has been proposed, which is usually referred to as \emph{Super-Humeanism} (see \citealp{485}, for the standard textbook on the subject). The Super-Humean metaphysics is relationalist, structuralist, and parsimonious in the same vein as ours, but it differs in some essential respects. First of all, Super-Humeanism is detached from any particular theoretical framework. Indeed, the main Super-Humean tenet is that the laws of any physical theory can be shown to supervene on a mosaic of permanent material points related by ever-changing distance relations. The fact that the fundamental picture of reality proposed by Super-Humeanism does not depend on physics is worrisome for many metaphysicians because it seems to deprive fundamental physics of much of its explanatory power (see \citealp{683}, for an argument in this sense). Indeed, Super-Humeanism seems to imply that physics provides no key insight into the nature of reality. Consequently, physics is threatened to be demoted to some sort of formal machinery that succinctly describes the very complicated motions of fundamental material particles. 

On the other hand, our metaphysics does depend on the particular physical domain considered for the characterization of both relations and relata making up a self-subsisting structure. Hence, as already pointed out, a ``Newtonian'' self-subsisting structure is different from a ``dynamical geometry'' one in many ontological respects that also have observational consequences. There is, of course, an important sense in which some structural aspects of the framework are model-independent, but this sense does not deprive physics of any explanatory power. On the contrary, it accounts for why some empirically adequate concepts are carried over in the development of physics. In other words, our metaphysics is indeed sensitive to the development of physics, to the point that we expect to radically revise the notion of self-subsisting structure at the quantum-gravitational regime (see \S \ref{sec:disc}). This is enough to defuse the charge of adopting what \citet[][p. 81]{684} calls ``Super-Humean subterfuge'':

\begin{quote}
Our physical description of the world exhibits the feature $X$ because the contingent relational distribution of matter throughout the history of the universe happens to be such that the best system description exhibits the feature $X$.
\end{quote}

As already pointed out, there are central features of our metaphysics that are not related to the mosaic in the way mentioned above but are instead entirely determined by the physics---e.g., the Euclidean nature of the spatial relations making up a Newtonian self-subsisting structure.


Another essential difference between our metaphysics and Super-Humeanism resides in the nature of the spatial relations postulated. For the Super-Humeans, the distance relations making up a configuration of matter points carry an extremely weak spatial connotation, which boils down to fulfilling the triangle inequality (see \citealp{485}, p. 22, for a list of requirements that any numerical assignment that coordinatizes a configuration has to obey). This, in turn, makes it highly improbable for a very large configuration of particles---like the one making up our universe---to be most simply embeddable in a space as low dimensional as a $3$-dimensional Riemannian manifold. Indeed, this would require such a configuration to obey by sheer chance an extremely high number of geometric constraints (see, again, \citealp{684}, p. 82, for a rough estimate of how improbable this might be). On the other hand, our self-subsisting structures do not fall prey to this kind of ``un-typicality'' argument. This is because they retain a much more robust geometric structure---mathematically represented by a conformal structure---which makes them straightforwardly embeddable in a low dimensional space (e.g., in the case of classical and relativistic motions). All we have to do to perform this embedding is to ``reverse'' the quotienting out procedure, thus getting back the starting embedding space (recall the discussion of the ephemeris equations in \S \ref{sec:ephs}). Of course, our reliance on a primitive conformal structure renders our metaphysics less parsimonious than the Super-Humean one, but we are more than happy to pay this price if, in return, we get a more plausible mechanism that accounts for how we get from self-subsisting structures to ordinary space.

\subsection{A Supervenience Basis for Space and Time}\label{sec:spst}

The above discussion clarifies the sense in which ordinary space supervenes on self-subsisting structures---at least, in the non-quantum case. In this picture, space is reduced in the strong mathematical sense encoded in the quotienting out procedure to the spatial relations making up a self-subsisting structure. This is how our metaphysics conforms to the Leibnizian motto that ``space is the order of co-existing things.'' Likewise, it is straightforward to recover standard space from a self-subsisting structure. This is done by reversing the quotienting out procedure that sets the shape space of the system under scrutiny, thus getting the standard picture where the \emph{relata} in the structure are now individual objects occupying a place in ordinary space (or field magnitudes making up a $3$-dimensional geometry in the dynamical geometry case). We have seen a concrete example where such a reverse procedure is implemented, i.e., the construction---which makes use of \eqref{ephemerisScale}---of an ephemeris scale from the purely intrinsic properties of a dynamical curve in shape space. In this case, the construction makes use of the shape potential, which can be reduced to facts regarding the structure of shapes.

However, this markedly reductionist attitude seems to face troubles when the task becomes making sense of the notions of time and change commonly used---especially in scientific practice. In a nutshell, we can put the challenge in the following way: How are we to recover richly structured notions like that of an arrow of time from the very faint-structured mosaic on which \eqref{curve0} supervenes? This is a particularly delicate question, given that the weak topological ordering among configurations postulated at the fundamental level makes it \emph{prima facie} implausible to recover a picture of time as an ``abstraction from change''---borrowing the famous characterization given by Ernst Mach. As we shall see in a moment, meeting this challenge involves some conceptual work, but it is by no means unfeasible. The strategy we will employ is similar to the spatial case, and amounts to showing how facts about time and change supervene on---in the strong sense of ``are reducible to''---the minimal set of fundamental relational facts inhering into the PSD mosaic. More precisely, we will couch the construction in terms of layers of description: The fundamental basis (i.e., the level-$0$ domain) featuring relational facts will be used to construct progressively more complex temporal notions. In order to keep the paper at a reasonable length, we cannot delve too deep into the conceptual subtleties involved in the construction. However, we submit that the sketch we are going to offer is sufficient to establish the feasibility and consistency of the program.

Before going through this construction, a remark is in place. One may wonder why, in our framework, recovering temporal notions should be more tricky than recovering spatial ones. The answer is quite easy: Because the framework accords a privileged ontological status to space as opposed to time. More precisely, self-subsisting structures possess a fundamental---albeit weak---spatial connotation in terms of the conformal structure inhering into them. From this point of view, as already said, recovering ordinary (instantaneous) space from a self-subsisting structure $\mathfrak{C}$ just amounts to embedding $\mathfrak{C}$ in the space that conveys the physical information encoded in $\mathfrak{C}$ in the simplest and strongest way. In this sense, via the embedding procedure, a self-subsisting structure becomes a ``universal snapshot'' of, say, a Newtonian $N$-particle world. Clearly, such a straightforward construction is not available in the temporal case. For starters, the embedding story should be applied to the sequence of self-subsisting structures constituting a possible world. However, we already know that this fundamental level---represented by an unparametrized curve in shape space---has no temporal connotation whatsoever, so some additional consideration has to be added on top of the geometric construction in order to show that some sort of temporal structure ``appears'' by virtue of carrying out said construction. Secondly, even granted that the timelike part of spacetime can be recovered in this way, still this construction would not account for the ``appearance'' of temporal notions that are not encoded in the spacetime picture (case in point, the notion of temporal passage). This remark justifies why, in our discussion, we are placing a particular emphasis on how time and change can be recovered in the PSD framework.

Let us now return to the discussion on the supervenience of temporal concepts in terms of levels of description. As already said, the level-$0$ domain is the mosaic itself. At this level, all the configurations making up a possible world---represented by a curve fulfilling \eqref{curve0}---are just given at once as a brute fact of nature. This means that the distinctness of such configurations is assumed as a primitive concept. This level-$0$ domain features fundamental facts regarding the ``nearness'' or ``similarity'' of neighboring configurations, making it the case that an unparametrized curve can describe these configurations in shape space. Facts regarding the similarity of neighboring configurations are also given as primitive. Remember from the discussion in \S \ref{subsec:structure} that this fundamental level just encodes facts about how self-subsisting structures are locally (i.e., topologically) ordered, which in particular means that there is no fact about a preferred direction of such ordering. Thus, this undirected ordering of self-subsisting structures resembles what, in the standard debate on the metaphysics of time, is known as a ``C-series'' ordering (this concept was famously introduced in \citealp[][p.~462]{6}; see also \citealp[][\S 3]{sep-mctaggart}, for a nice historical discussion of McTaggart's argument against the reality of time). To be fair, this resemblance should be taken \emph{cum grano salis} since, in the standard debate, time series are usually understood as orderings among events or ordinary facts, not (facts about) physical structures in the sense we adopt.

From this set of primitive facts obtaining at a world, more structured notions can be shown to supervene. Thus, the level-$1$ domain involves a global ordering over the set of configurations actualized at a world. This global ordering is reminiscent of the ``B-series'' ordering in terms of an ``earlier than'' relation. Such an ordering is required to make sense of ordinary change, but it is not ``there'' in the mosaic, i.e., it is not a fundamental feature of reality; rather, it can be shown to supervene on the mosaic through mathematical reduction. Indeed, this level-$1$ ordering can be constructed by arbitrarily fixing a reference configuration and then by assigning a ``numerical flag'' to the neighboring configurations based on facts regarding their degree of similarity to the starting one. In this way, the intrinsic structure of shapes is exploited to establish an ordering that can be described in terms of an arc-length parametrization of the curve (with respect to, e.g., the kinematic metric on shape space). Thanks to this further structure, we can traverse the whole curve as the parameter values increase---or decrease---and from this recover the usual talk of spatial relations ``changing'' from one configuration to the other as the curve is traversed. Note how this connotation of change is not primitive but given in terms of primitive distinctness and similarity.

This setting also permits---at least in well-behaved cases---to establish a notion of ``identity over configurations'' for both relations and \emph{relata}. Take, for example, the case of the Newtonian $N$-particle system. In this case, each shape consists of $N$ particles related through Euclidean spatial relations but, according to the metaphysics we put forward, there is no fundamental fact of the matter about the identity of the particles and relations making up a shape being carried over to the other configurations: Each self-subsisting structure has ``its own'' particles and relations. By using the arc-length parametrization to traverse the curve, we automatically relate particles and relations in subsequent configurations by establishing that moving from one to the other represents the change in the spatial configuration of a \emph{unique} set of $N$ particles linked through Euclidean spatial relations. Pictorially speaking, this is similar to animating a picture by quickly showing a series of frames ordered in terms of their similarity. Obviously, each frame shows a picture different from the others, but by showing them in the specific order induced by their similarity, we gain the impression of a unique subject moving (i.e., changing its spatial configuration in time). One may object that this animation analogy sneaks in some decidedly temporal notion in the discussion: After all, animations unfold \emph{in time}. Our reply is that this objection gets things backwards, so to speak: It assumes that time is required to make sense of change, whereas our construction is exactly meant to show that nothing over and above a global and directed ordering among self-subsisting structures is needed to make sense of change as a succession of ``frames.'' As \citet[][p.~163]{sav} puts it: ``We do not need an animated picture to have a picture of animation.''

Furthermore, by establishing a measure of ``how much structure'' a configuration contains---as represented, in well-behaved Newtonian cases, by the complexity function \eqref{complexity}---it is moreover possible to establish a notion of duration. This notion is arrived at by constructing a temporal metric that, in the non-relativistic case, gives precisely the Newtonian time $t$ (recall the construction of the ephemeris duration equation \eqref{ephemerisDuration} sketched in \S \ref{sec:ephs}). This is the level-$2$ domain of time, which accounts for the physical appearance of clocks marching in step. Also in this case, it is worth noting that there are no fundamental facts about a global metric of time crafted in the mosaic but, instead, facts about duration supervene on primitive facts about shapes' complexity through an arbitrary---albeit natural, in many cases---mathematical construction such as that of equation \eqref{ephemerisDuration}. The construction of this second layer is based on the dynamical formation of stable records. The assessment of the stability of such records---which are nothing but clusters of particles in the classical case---is made possible by invoking the notion of identity over configurations that comes at the level-$1$ domain. So, for example, the talk of clocks in the Newtonian case turns out to be a concise yet informative way to referring to the dynamics of the Kepler pairs mentioned at the end of \S \ref{dynamics}.

Finally, the level-$3$ domain of time, that is, the appearance of a directed flow of time, is accounted for in terms of the secular growth of the complexity function \eqref{complexity}. More concretely, we can establish a notion of directed temporal passage by individuating a certain starting self-subsisting structure and then ``trace'' its development through the dynamical curve (again, by exploiting level-$1$ notions). Simply speaking, this ``present'' configuration moves away from the past and into the future following the secular growth of complexity inherent into the curve. This is how an ``A-series'' of time is recovered in this context. Of course, such a construction is entirely arbitrary---in the sense that there is no genuine fact about something being past, present, or future---and is meant to show how a tensed description of a physical process can be achieved in our framework. 

The layered structure sketched above can be summarized using the following diagram:

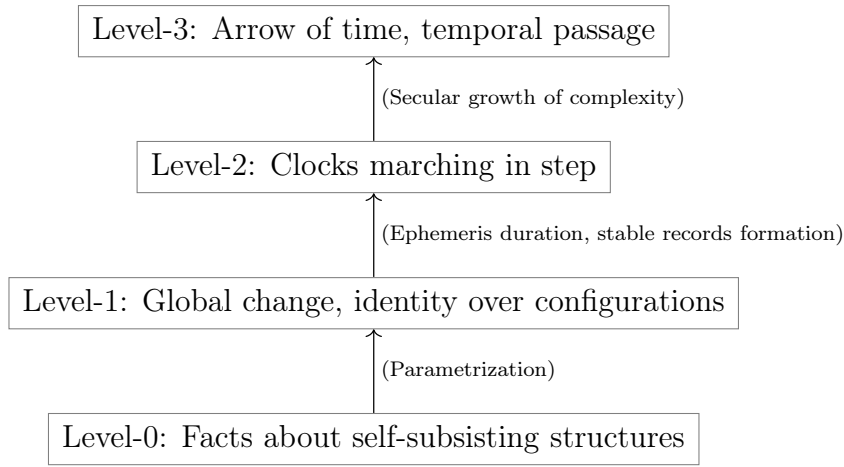
\begin{center}
\begin{tikzcd}[row sep=large, cells={nodes={draw=gray}}]
\text{Level-$3$: Arrow of time, temporal passage}\\
\text{Level-$2$: Clocks marching in step}\arrow[u, "(\text{Secular growth of complexity})"']\\
\text{Level-$1$: Global change, identity over configurations}\arrow[u, "(\text{Ephemeris duration}\text{, stable records formation})"']\\
\text{Level-$0$: Facts about self-subsisting structures} \arrow[u, "\text{(Parametrization)}"']
\end{tikzcd}
\end{center}

As already stressed several times, only the bottom level of the diagram is concerned with the ontology; the subsequent levels deal with the description of such underlying reality. Also, note that this reconstruction of change, duration, and an arrow of time is meant to show how the level-$0$ mosaic grounds these notions as used in physics (and in scientific practice in general). Of course, a more detailed story should be provided as to how we get to \emph{perceive} change and the directed passing of time. This task goes far beyond the scope of the present paper and involves explaining how minds emerge as subsystems of self-subsisting structures and how they get correlated with the fundamental facts grounding the higher-level temporal notions characterized above. This is undoubtedly an interesting future line of research. 

To conclude this section, we point out that there can be possible worlds in which the amount of structure formation inherent into the mosaic is not enough to ground the emergence of level-$3$, or even  level-$2$ temporal structures: Think, for example, of a universe eternally in thermal equilibrium. In such worlds, it makes no sense to speak of clocks, let alone a preferred direction for the passing of time.

\section{Conclusion: Pure Shape Dynamics in Perspective}\label{sec:disc}

PSD represents a natural evolution of the relational framework originally put forward by Barbour and Bertotti. It promises to deliver a complete geometrization of the dynamics, which eschews from the physical picture any remnant of non-intrinsic characterization of the evolution of a system. In this sense, the most remarkable feature of PSD is that it does not need any notion of parametrization to make sense of the dynamics. We can hence talk of \emph{strong} temporal relationalism, as opposed to the weaker type that does require a notion of monotonically increasing parameter to establish a way in which the dynamics unfolds.

Furthermore, the PSD's framework is all-encompassing in that it can, in principle, accommodate relational counterparts of physical models pertaining to classical, relativistic, and quantum physics (this latter case being the focus of a companion paper). Such inter-theory relation between standard and relational models is rendered in terms of a ``quotienting out'' procedure that eliminates all the non-relational degrees of freedom from the former models, thus generating the relational ones of PSD.

From a metaphysical perspective, we have argued that the PSD framework naturally goes hand in hand with an ontic structural realist and Humean take on the laws of physics, which means that PSD is a theory of self-subsisting structures---i.e., structures characterized wholly intrinsically. As we have pointed out before, however, structuralism alone is not enough to fully establish the ontological nature of the relations and \emph{relata} entering a model of PSD. Both relations and \emph{relata} then have to be specified via a reconceptualization of the ontology of the starting non-relational models. This need for a reconceptualization implies that physics does have a say in our metaphysical framework, differently from the Super-Humean approach that fixes \emph{a priori} the ontology of physics. The clearest example of reconceptualization is represented by the case of material particles moving in a background space in classical mechanics, which, in our structuralist and Humean reading of PSD, become a Leibnizian structure with identityless and propertyless \emph{relata} individuated solely through the Euclidean spatial relations they stand in. 

We have also clarified that the models of PSD should be taken at face value only insofar as their (empirically adequate) relational aspects are concerned, thus considering their complete ontological characterization as physical \emph{possibilia} rather than faithful pictures of how the actual world fundamentally is. This begs the question as to which sector of PSD we should then take to deliver the set of truly fundamental models, i.e., those that fully and adequately describe the actual world. An answer to this question depends on whether PSD may eventually succeed as a ``final'' physical theory. This represents a significant challenge to the future developments of the framework for many reasons, both technical and conceptual.

Let us start by pointing out that such an advanced sector of PSD is likely to account for the quantum \emph{and} gravitational aspects of the dynamics. Otherwise said, the quantum models of PSD and the dynamical geometry ones are expected to be some sort of low energy and long-range limits or approximations of these yet-to-be worked out fundamental models of the theory. One may be then tempted to classify this fundamental sector of the theory as the ``quantum-gravitational'' one. However, that would be misleading because, in PSD, there are no genuinely ``quantum'' or ``gravitational'' features of the physical description of the world: Everything is reduced to geometric degrees of freedom of dynamical curves on shape space (which in turn are just a simple yet informative way to refer to the mosaic made of self-subsisting structures). Indeed, the companion paper will discuss the possibility of ``geometrizing away'' the wave function of a system and its associated transition amplitudes. If such a radical geometrization of physics is viable, there will then be a clear sense in which this hypothetical fundamental sector of the theory represents a pre-quantum-gravitational reality from which both the quantum and the gravitational aspects of physics appear in the appropriate limit or approximation. This highlights how the PSD framework is a potentially novel approach to quantum gravity that dispenses with well-known issues related to the quantization of the gravitational field.

The above remark shows how formidable it is the task of working out such a fundamental sector of PSD. Indeed, while the non-fundamental sectors of the framework can be constructed as quotiented out versions of existing non-relational theories, such a strategy is not available in the fundamental case for two reasons. First, there is no ``final'' theory of quantum gravity as yet and, second, it would be improbable that such a theory comprised spatiotemporal background structures (to the contrary, such a theory should exactly be about how it comes that background independence is a fundamental feature of reality). Hence, the development of this sector of PSD should be carried out ``on its own,'' so to speak. This challenge opens up a plethora of conceptual questions, one of the most pressing being: What would become of self-subsisting structures in such a fundamental regime if, as it is very likely, no spatial or also quantum notions were available? In other words, what would it be the relation making up these structures if they were not accounted for in terms of spatial or even entanglement relations? Furthermore, what would make these relations \emph{physical} as opposed to merely mathematical constructions if they could not be characterized in any of the above ways?

In short, the PSD framework represents a largely unexplored territory that is full of promises but, at the same time, gives no guarantees as yet of total success. However, in the present paper, we hope to have convincingly shown that the PSD project---in both its physical and metaphysical aspects---is worth pursuing because it provides new perspectives on current physics and promises to deliver a novel take on the problem of quantum gravity.

\pdfbookmark[1]{Acknowledgements}{acknowledgements}
\begin{center}
\textbf{Acknowledgements}
\end{center}
We thank an anonymous reviewer for the very helpful comments on an earlier version of this manuscript. AV and PN gratefully acknowledge financial support from the Polish National Science Centre, grant No. 2019/33/B/HS1/01772.

\bibliography{biblio}

\end{document}